# Mechanism of Spin Scattering in Ta investigated by Scanning Inverse Spin Hall Effect Meters

Peiwen Luo[1], Zhe Wu[1], Fei Huang[2], Bing Peng[2], Wenxu Zhang [2,a)]

1. School of Physics, University of Electronic Science and Technology of China, Chengdu 611731;
2. State Key Laboratory of Electronic Thin Films and Integrated Devices, University of Electronic Science and Technology of China, Chengdu 611731

**ABSTRACT:** In this work, a scanning inverse spin Hall effect measurement system based on a shorted coaxial resonator has been built, which provides a high throughput method to characterize spin transport properties. The spin diffusion length of Ta at room temperature is determined via automatic measurements of Py/Ta bilayer strips with different thicknesses of Ta. The results show that the spin diffusion length is about 4 nm with conductivity about $7.5 \times 10^5$ $\Omega^{-1}m^{-1}$, which lead to the conclusion that the intrinsic mechanism of spin relaxation of Ta is the Elliott-Yafet interactions. The setup developed in this work provides a convenient, efficient and nondestructive way to obtain the spin and electron transportation characteristics of the spintronic materials, which will fertilize this community by developing new materials and figuring out their mechanism.

Spin pumping (SP), which transports polarized electron spins from a ferromagnet into the adjacent normal metal layer with assistance of microwaves, is one of the most efficient method [1,2] for spin injection. It can avoid the conductance mismatch [3,4] so that the spins can be injected from metals to semiconductors or vice visa. The injected spins are in nonequilibrium and can convert into charge currents during diffusion in the materials due to the spin orbital coupling, which is commonly known as the inverse spin Hall effect (ISHE)[1]. The coupling effects are essential to develop spintronic devices. There is one crucial parameter called spin diffusion length (λ), which measures the distance over which electron spins flipped [2]. There are several ways to determine this parameter, such as Andreev spectroscopy [3], spin magnetoresistance [4], spin value [5], ISHE [6], and non-local method [7]. In the non-local method, lithography with precision of nanometer is quite demanding. Among them, the SP-ISHE effect is widely used due to the high spin injection efficiency and low demanding of lithography. The experiments are usually measured by a resonator [8], coplanar waveguide [9] or transmission line [10]. Tens of samples with different thickness are prepared and measured separately [8,11], the spin diffusion length is determined by fitting the data with different film thickness [8].

As proposed by Meinert et al. terahertz emission spectroscopy and broadband ferromagnetic resonance are the two lithography-free high-throughput ways to measure the ISHE [11]. In this letter, we proposed a high through-put way to measure the ISHE with a shorted microwave probe. The shorted probe is already used to characterize the local ferromagnetic resonance (FMR) properties in a broadband frequency range, where the permeability mapping can be obtained [12,13]. In our experiments, the samples are prepared on a wafer and patterned with lithography or rigid masks. The measurements can be automatically performed with the help of motorized stage. By this

a) Electronic mail: xwzhang@uestc.edu.cn

method, several parameters such as conductance, anisotropic magnetoresistance, FMR properties, photo-voltage from spin pumping can be measured simultaneously.

The shorted microwave probe used in this experiment is a semirigid coaxial cable with length of 50 mm, which is terminated with a gold foil with width of 2 mm at one end, as shown in Fig. 1(a). The diameters of the inner and outer conductor of the probe are 3 mm and 6.35 mm, respectively. When the shorted probe is coupled to a RF generator, a microwave magnetic field in the **y** direction will be generated by the microwave current in the gold foil at the shorted-end. The magnitude of $\mathbf{h}_y$ is about 20 A/m at 4 GHz according to the simulations [inset of Fig. 1(b)] with the Finite Element Method. The S-parameter $S_{11}$ of the probe [Fig. 1(b)] was obtained by a VNA (Agilent N5234A).

Stripes with permalloy/non-ferromagnetic metal (Py/Ta) bilayers are used to detect the ISHE voltages converted from the spins injected by the spin pumping. The geometries for the ISHE detection of the spin current is shown in Fig. 1(c). We use a Py/Ta bilayer film with interfaces normal to *z*. The width and the length of the sample are fixed to be 40 μm and 8 mm, respectively, to suppress the spin rectification effect (SRE) according to our previous work [14]. The permalloy films are sputtered onto a thermally oxidized silicon wafer by magnetron sputtering. Two aluminum pads with 2×2 mm$^2$ were sputtered on both ends of the bilayer strips to enhance the electrical connection with the metallic spring probes, which are fixed at the opposite sides of the coaxial cable to measure the DC voltages generated during the spin pumping. The block diagram of the automatic SP-ISHE measurement system is drawn in Fig. 1(d). Modulated microwaves generated by a signal generator are fed into the shorted probe after amplification by a power amplifier. Meanwhile, a reference signal with same frequency as the modulated wave is sent to a lock-in amplifier. The DC magnetic field is provided by an electromagnet. The sample can move in three directions (XYZ) through the motorized stage, thus the voltage signals in different area of a wafer sample can be obtained.

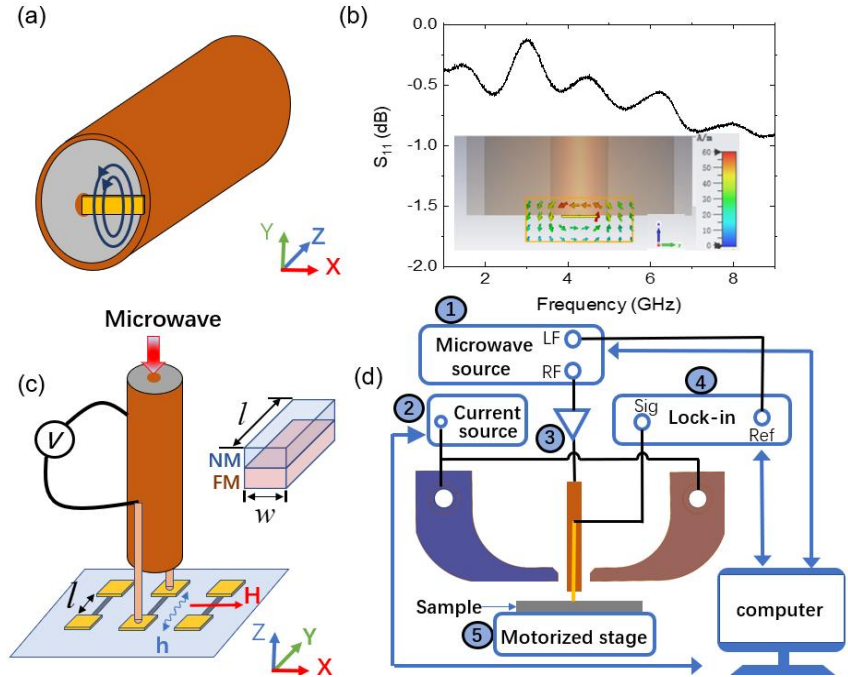

Fig. 1. (a) Schematic drawing of the shorted probe. (b) $S_{11}$ curve of non-load the shorted probe, the inset is the simulation result of microwave magnetic field. (c) Schematic drawing of the setup for measurement by the shorted

probe. The length *l* and width *w* of the bilayer are 8 mm and 40 μm, respectively. FM and NM represent the Py and Ta layers respectively. (d) Block diagram of the testing system. ① microwave source [Rohde&Schwarz, SMB 100A]; ② current source [Iteches, IT6432]; ③ home-made power amplifier; ④ lock-in amplifier [Stanford Research System, SR830]; ⑤ motorized stage [Thorlabs, LTS300].

According to the spin pumping model, the spin current density is proportional to the RF power ($P_{MW}$) in the linear regime [15]. Fig. 2(a) shows the microwave power $P_{MW}$ dependence of the ISHE voltage for the Py(20)/Ta(15) film, where the number in the parenthesis is the thickness of the film in nm. The result indicates that the microwave power (<250 mW) is lower than the saturation of the resonance absorption for the present system and the devices are working in the linear regime. In addition to that, the microwave magnetic field excited by the shorted probe is sensitive to the variation of the probe-to-sample distance (*d*). In order to optimize the *d* for the measurement, we measured the ISHE voltages with different *d* and the results are shown in Fig. 2(b). It can be seen that the $V_{ISHE}$ decreases exponentially with the increase of *d*. We fit the data with $V_{ISHE} = V_0 e^{-d/\tau_V}$, and get the decay constant $\tau_V = 0.45$ mm. Meanwhile, the amplitude of the microwave magnetic field $\mathbf{h}_y$ also decreases exponentially from the short-end of the probe, which can be modeled as $h_y^2 = h_{y0}^2 e^{-d/\tau_m}$. From the results by our FEM simulations [Fig. 2(c)], we get the decay constant $\tau_m = 0.47$ mm, which agrees with the decay of the ISHE voltages, confirming the linear relationship between them. We set the *d* to 50 μm and $P_{MW}$ to 250 mW in the following measurements to get a well controlled signal amplitude.

We first prepared 18 independent Py/Ta bilayer strips equidistance on a wafer, where strips are of the same geometries and the distance between each strip is 3 mm. The measured ISHE voltage with $P_{mw}$ = 250 mW is shown in Fig. 2(d), where the horizontal axis indicates the relative positions of the stripes. The photo of the samples is shown in the inset of Fig. 2(d). It can be seen that the ISHE voltages of film samples at different positions are basically the same. The average value and its standard deviation of the voltages is 1.20 μV and 0.026 μV, respectively, which sets the error bar of the voltages measured by our system.

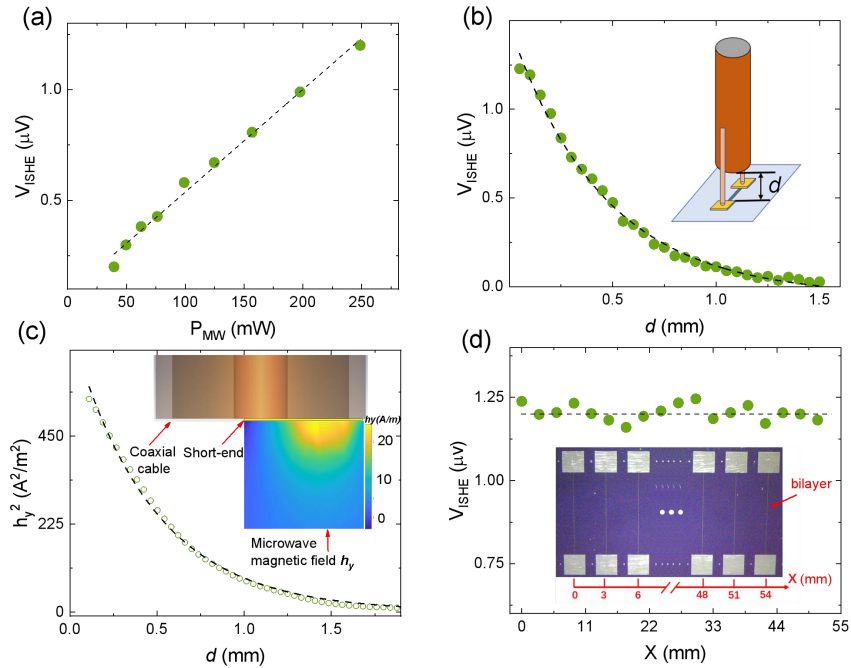

Fig. 2. (a) The microwave power dependence of the ISHE voltage. (b) The probe-sample distance (*d*) dependence

of the ISHE voltage for the Py/Ta bilayer film. (c) The simulation results of Y component of microwave magnetic field $h_y^2$ as a function of $d$. The inset is the simulated result of $h_y$ at the terminal of the short-end probe. (d) The position X dependence of the inverse spin Hall voltage ($V_{ISHE}$) measured for the strips under the 250 mW microwave excitation. The dashed line shows the average value of voltages. The inset is the photo of samples.

In the characterization of spin transportation, one of the key parameters is the spin diffusion length λ, which can be determined from the thickness dependent SP-ISHE voltages. According to the model derived by Ando et al [16], the relationship between the inverse spin Hall voltage $V_{ISHE}$ and the parameters of a FM/NM bilayer film is described as:

$$V_{ISHE} = \theta_{SH} \lambda j_s^0 \left(\frac{2e}{\hbar}\right) \frac{l}{d_N \sigma_N + d_F \sigma_F} \tanh\left(\frac{d_N}{2\lambda}\right) \quad (1)$$

where $\theta_{SH}$, $j_s^0$, $e$, $\hbar$, $\sigma_N$, $\sigma_F$, $d_N$, $d_F$ and $l$ denote the spin Hall angle of the NM layer, the spin-current density at the interface of FM/NM bilayer, the electron charge, the reduced Planck constant, the conductivity of NM layer, the conductivity of FM layer, the thickness of NM layer, the thickness of FM layer and the length of the sample. In order to measure the spin diffusion length, we deposit Ta layer with different thickness by placing a wafer substrate off the center of the magnetron gun. We use the same mask to pattern the film as the one used above in Fig. 2(d). The film thickness is determined by a separate measurement. We determine the conductivities of the NM ($\sigma_N$) and FM ($\sigma_F$) layers by fitting the conductance of the films with different thickness of Ta layer, which can be obtained as: $\sigma_F = 1.8 \times 10^6$ $\Omega^{-1}m^{-1}$, $\sigma_N = 7.56 \times 10^5$ $\Omega^{-1}m^{-1}$.

The SP-ISHE voltages and its dependence of the Ta thickness are shown in Fig. 3. The voltage curves are all symmetric Lorentzian functions of the applied field, which indicates that the antisymmetric ones originated from the SRE were suppressed[14]. This purified signal enables us to analyze the spin-charge conversion. We fit the measured data to Equ. (1) and show in the figure with a dashed curve. As can be seen from the fitting curve, the $V_{ISHE}$ increases with $d_N$ when $d_N$ is less than 10 nm. The $V_{ISHE}$ decays with increasing $d_F$ when $d_F$ is greater than that. This is due to the increment of pumped spin currents and the reduction of resistance with increasing $d_N$. The measured $V_{ISHE}$ values are well reproduced with the parameter λ= 4 nm, which is comparable to 4.8 nm obtained with CoFeB/Ta bilayers, where the conductivity for Ta is $7.45 \times 10^5$ $\Omega^{-1}cm^{-1}$.[9]

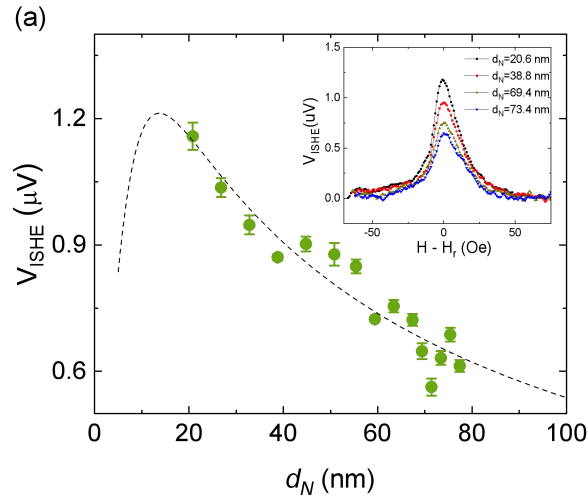

Fig. 3. Dependence of $V_{ISHE}$ on the thickness of the NM layer ($d_N$) measured for the Py/Ta bilayer films. The dashed line is a fit using Eqs (1). The inset shows the H dependence of $V_{ISHE}$ in bilayers with four selected $d_N$'s.

We finally give some discussions on the mechanism of the spin scattering. According to the model of spin diffusion length proposed by Valet and Fert[17], one has $\lambda = \sqrt{D\tau_s}$, where the diffusion constant $D$ depends on the Fermi velocity $v_F$ as $D = \tau_e v_F^2$. When the spin relaxation time $\tau_s$ is given by Elliott-Yafet model[18] it is proportional to the electron momentum relaxation time $\tau_e$. We thus have $\lambda \propto \tau_e$. The electron conductivity $\sigma = \frac{ne^2}{m}\tau_e$ according to the Drude model, where $n$ is the electron density. Thus we conclude that $\lambda$ is proportional to $\sigma$. We compare the spin diffusion length of Ta with different conductivity collected from previous publications as shown in Fig. 5, a linear relationship between them is clear, which points to the conclusion of the relaxation of spin by the Elliott-Yafet mechanism.

We also notice that the interface between the FM and NM contributes significantly to the spin transportation when the film is thin [4, 9], where the conductivity will decrease due to the interface scattering. The Rashba effect can give rise to the spin flip scattering, which introduces D'yakonov-Perel' mechanism into the play that gives $\tau_s$ proportional to $\tau_e^{-1}$.

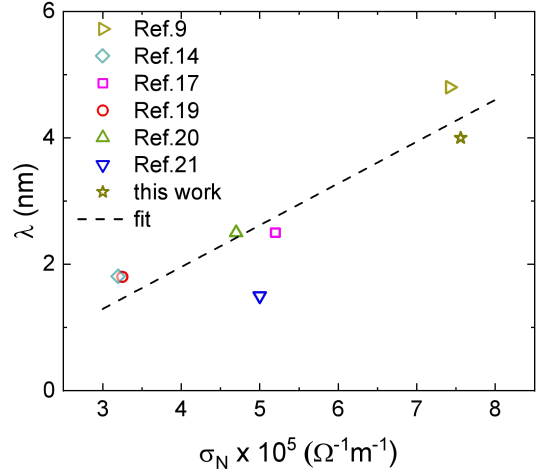

Fig. 4. Spin diffusion length (λ) of Ta as a function of conductivity ($\sigma_N$). The dashed line is a linear fit of the data for guiding eyes.

We see the complications in the spin-charge conversion processes. The scanning ISHE system proposed in this work provides a unique way to realize simultaneous measurement of samples prepared in the same run, so that the parameters which influence the properties can be more systematically controlled. At the same time, electron and spin transportation properties can also be measured nondestructively, which can benefit discussions of the relationships between them.

In summary, we have constructed a scanning ISHE test system using a shorted microwave probe. The major advantage of the system is that it can measure different samples on the same wafer with the scanning method, providing a new solution for the characterization of spin transport properties of materials. The Py/Ta bilayer strips with different Ta thickness $d_N$ were measured. From the $d_N$ dependence of bilayer film conductance, the conductivities of Ta and Py are fitted to be 7.56 × 10⁵ Ω⁻¹m⁻¹ and 1.8 × 10⁶ Ω⁻¹m⁻¹, respectively, together with the spin diffusion length of Ta λ = 4 nm from fitting the SP-ISHE voltages. We infer that the intrinsic mechanism of Ta is Elliot-Yafet scattering from the linear relationship between the conductivity and spin diffusion length of this work and other literatures. The system provides a high throughput method to characterize spintronic materials, where the spin and electron transportation properties are obtained automatically and nondestructively.

**Acknowledgements**

Financial support from the National Natural Science Foundation of China (Grant No. 12274060) was greatly acknowledged.